\documentclass[prl, reprint, superscriptaddress]{revtex4-2}
\usepackage{amsmath}
\usepackage{amssymb}
\usepackage{float}
\usepackage{graphicx}
\usepackage[caption=false, position=top, singlelinecheck=off, justification=raggedright]{subfig}
\usepackage{multirow}
\usepackage{color}
\usepackage{pst-node}
\usepackage[unicode]{hyperref}
\hypersetup{
	unicode=true,         	
	colorlinks=true,      	
	linkcolor=blue,		   	
	citecolor=blue,       	
	urlcolor=blue		   	
}

\def\maketitle{ %
 \@author@finish
\title@column\titleblock@produce
\suppressfloats[t]%
 \let\and\relax
\let\affiliation\@gobble@opt@one
 \let\author\@gobble
 \@author@init
 \let\@authors\@empty
\let\@authors@curr\@empty
\let\@affil@list\@empty
 \let\keywords\@gobble
 \let\@keywords\@empty
 \let\email\@gobble
 \let\@address\@empty
 \let\maketitle\relax
 \let\thanks\@gobble
 \titlepage@sw{ %
 \clearpage
}{}%
 }%

\begin{document}

\makeatletter
\def\maketitle{
\@author@finish
\title@column\titleblock@produce
\suppressfloats[t]}
\makeatother

\title{Vortex and soliton dynamics in particle-hole symmetric superfluids}

\author{Jim Skulte}
\affiliation{Zentrum f\"ur Optische Quantentechnologien and Institut f\"ur Laserphysik, 
	Universit\"at Hamburg, 22761 Hamburg, Germany}
\affiliation{The Hamburg Centre for Ultrafast Imaging, Luruper Chaussee 149, 22761 Hamburg, Germany}

\author{Lukas Broers}
\affiliation{Zentrum f\"ur Optische Quantentechnologien and Institut f\"ur Laserphysik, 
	Universit\"at Hamburg, 22761 Hamburg, Germany}

\author{Jayson G. Cosme}
\affiliation{National Institute of Physics, University of the Philippines, Diliman, Quezon City 1101, Philippines}

\author{Ludwig Mathey}
\affiliation{Zentrum f\"ur Optische Quantentechnologien and Institut f\"ur Laserphysik, 
	Universit\"at Hamburg, 22761 Hamburg, Germany}
\affiliation{The Hamburg Centre for Ultrafast Imaging, Luruper Chaussee 149, 22761 Hamburg, Germany}

\date{\today}
\begin{abstract}
We propose to induce topological defects in particle-hole symmetric superfluids, with the prime example of the BCS state of ultracold atoms and detect their time evolution and decay. We demonstrate that the time evolution is qualitatively distinct for particle-hole symmetric superfluids, and point out that the dynamics of topological defects is strongly modified in particle-hole symmetric fluids. We obtain results for different charges and compare them with the standard Gross-Pitaevskii prediction for Bose-Einstein condensates. We highlight the observable signatures of the particle-hole symmetry in the dynamics of decaying solitons and subsequent vortices.
\end{abstract}
\maketitle
The presence or absence of particle-hole symmetry in a physical system is a fundamental property pervading its dynamical properties. Particle-hole symmetry is realised in Lorentz invariant theories such as the standard model of elementary physics \cite{Weinberg}, low energy effective models close to  quantum criticality \cite{sachdev_2011}, and the famous Bardeen-Cooper-Schrieffer (BCS) theory of superconductivity \cite{Pekker_2015,Varma}, note \footnote{ Due to the close connection between relativistic Lorentz invariance and particle-hole symmetry, models that are particle-hole symmetric are also sometimes refereed to as relativistic models.}. We note that the order parameter dynamics of high-$T_\mathrm{c}$ superconductors can be described by an effective particle-hole symmetric theory, which allows for exploring the dynamics of the Higgs/amplitude mode \cite{Homann2020,homann2020higgsmediated,dai2021photoinduced,dai2021}. Similarly, in ultracold neutral atoms the emergence of an effective particle-hole symmetry has been theoretically predicted \cite{Altman2002,Pollet2012} and confirmed experimentally  \cite{Leonard,Endres_2012}. Recently, amplitude oscillations of the order parameter in the BEC-BCS crossover have been reported \cite{Behrle}, suggesting the presence of approximate particle-hole symmetry. \\
\indent The dynamics of topological defects, such as solitons and quantized vortices, derives from and exemplifies the properties of the underlying quantum fluid. The stability of solitons  has been discussed extensively for the non-linear Schrödinger equation or Gross-Pitaevskii equation (GP) \cite{Zakharov,Jones1986,Brand,Mateo,Kevrekidis}.  Zakharov and Rubenchik coined the term 'snaking` to refer to the characteristic bending of solitons prior to their decay. Snaking is a manifestation of the Magnus force. This has been discussed for neutral bosonic systems within the GP equation \cite{Guenther,Toikka_2017,Adams,Holland,Sheehy,Denschlag97,Sengstock} , in the BEC-BCS crossover \cite{Zwierlein2,Zwierlein,Tempere2019}
, as well as in superconductors \cite{Bardeen,Thouless}.\\
 \indent We propose
\begin{figure}[t!]
\label{fig:1}
\centering
\includegraphics[scale=0.27]{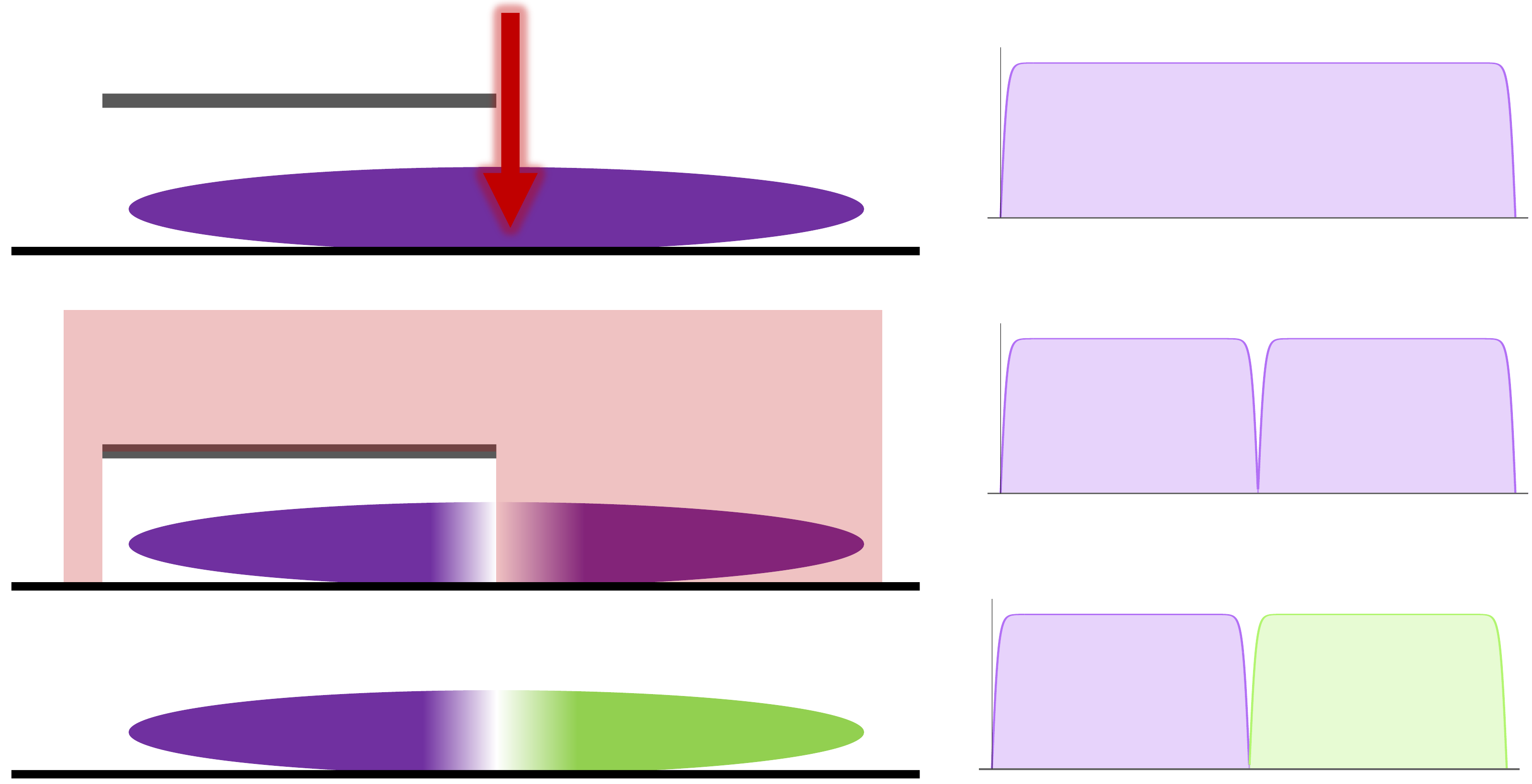}
\begin{picture}(380,5)
\put(-10,137){\makebox(20,0){\color{black}(a)}}
\put(-10,92){\makebox(20,0){\color{black}(c)}}
\put(-10,37){\makebox(20,0){\color{black}(e)}}
\put(142,137){\makebox(20,0){\color{black}(b)}}
\put(142,92){\makebox(20,0){\color{black}(d)}}
\put(142,47){\makebox(20,0){\color{black}(f)}}
\put(50,104){\makebox(20,0){\color{white}$\theta=0$}}
\put(95,104){\makebox(20,0){\color{white}$\theta=0$}}
\put(50,50){\makebox(20,0){\color{white}$\theta=0$}}
\put(95,50){\makebox(20,0){\color{white}$\theta=0$}}
\put(50,19){\makebox(20,0){\color{white}$\theta=0$}}
\put(95,19){\makebox(20,0){\color{black}$\theta=\pi$}}
\put(144,125){\makebox(20,0){\color{black}$|\psi|^2$}}
\put(190,113){\makebox(20,0){\color{black}$\theta=0$}}
\put(144,82){\makebox(20,0){\color{black}$|\psi|^2$}}
\put(172,70){\makebox(20,0){\color{black}$\theta=0$}}
\put(215,70){\makebox(20,0){\color{black}$\theta=0$}}
\put(144,37){\makebox(20,0){\color{black}$|\psi|^2$}}
\put(172,25){\makebox(20,0){\color{black}$\theta=0$}}
\put(215,25){\makebox(20,0){\color{black}$\theta=\pi$}}
\put(76,45){\makebox(10,10){\color{black}$\leftrightarrow$}}
\put(76,52){\makebox(10,10){\color{black}$2\xi$}}
\put(76,15){\makebox(10,10){\color{black}$\leftrightarrow$}}
\put(76,22){\makebox(10,10){\color{black}$2\xi$}}
\end{picture}
\label{fig:1}
\caption{Schematic representation of the proposed protocol to imprint a soliton in (a),(c),(e) and the corresponding density profiles $|\psi|^2$ and phase distributions $\theta$ in (b),(d),(f). A narrow laser sheet is applied to a quantum fluid on the BEC side of the crossover to create a density depletion in the condensate (a). The quantum fluid is split into two subsystems with a relative phase of zero, see (b) and (d) .  Next, a $\pi$-pulse is applied to half of the condensate (c) to create the phase pattern of a dark soliton, see (e) and (f). Next, the interaction is adiabatically changed across the crossover deep into the BCS side. The narrow laser sheet separating the two subsystems is removed, which triggers the soliton dynamics.  $\xi$ is the healing length of the condensate. }
\label{fig:1}
\end{figure}
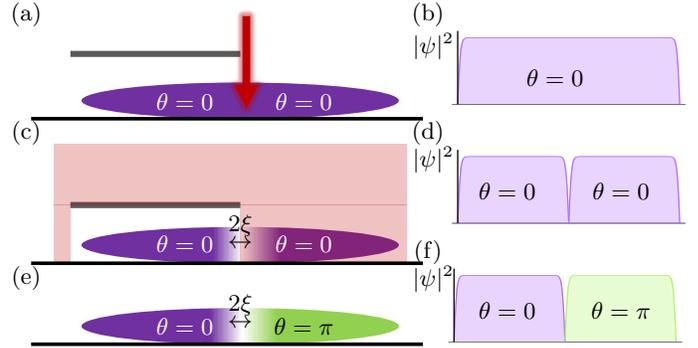  
  to determine the influence of particle-hole symmetry on the dynamics of topological defects in  two-dimensional (2D) neutral superfluids. We focus on the BCS state as our primary example, but our results hold for any approximately particle-hole symmetric system, e.g. bosons in an optical lattice near unit filling \cite{Endres_2012}. For this purpose we present the similarities and differences in the dynamics of topological defects in the absence and presence of particle-hole symmetry. We also compare the dynamics of the particle-hole symmetric theory for zero and nonzero Noether charge, corresponding to a balanced mixture of particles and holes and an imbalanced mixture of particles and holes, respectively. We find that the case with nonzero charge is reminiscent of the dynamics of the GP equation. On the other hand, for vanishing charge, in which the number of particles and holes is balanced, we show that vortices do not experience any Magnus force. This leads to a soliton decay without snaking, setting it apart from soliton dynamics in non-particle-hole symmetric fluids, such as BECs.  To induce soliton dynamics of the quantum fluid in the BCS limit, we propose to imprint a soliton on the BEC side of the crossover in the presence of a potential barrier. As a next step, we propose to ramp the fluid adiabatically across the crossover into the BCS limit, while keeping the barrier potential up. Finally, the barrier potential is ramped to zero, to induce the soliton dynamics. This protocol of initializing the dynamics enables imprinting of the phase pattern with an off-resonant optical pulse, whereas direct phase imprinting in the particle-hole symmetric limit is prohibited. We note that this statement holds only for an exact particle-hole symmetric case. In experiments such as those in Ref.~\cite{Zwierlein}, particle-hole symmetry is only approximately realised. That is, the appropriate effective action is expected to have both $K_1 \partial_t$ and $K_2 \partial^2_t$ contributions, as we discuss below. The $K_1 \partial_t$ term allows the phase imprinting as it is the dominant term in the BEC regime. The proposed protocol is displayed in Fig~\ref{fig:1}. \\ \indent
We consider a low-energy effective models of the form \cite{Pekker_2015}
\begin{align}
\label{action}
\mathcal{S}= \int d^2x dt \left\{ K_2 \left(\partial_t \psi\right)\left( \partial_t \bar{\psi}\right) -iK_1\left( \partial_t \psi \right) \bar{\psi}-\frac{1}{m}\nabla \bar{\psi} \nabla \psi  \right. \notag \\ \left. -\mu|\psi|^2+\frac{g}{2}|\psi|^4+V_\mathrm{ext}|\psi|^2  -i \mu_\mathrm{Q}\left( (\partial_t \bar{\psi})\psi-\bar{\psi}\partial_t \psi \right)   \right\}~,
\end{align}
where $K_{1,2}$ are the above-mentioned parameters that determine the time dependence, $\mu$ is the square root of the gap energy, which has the dimensions of a mass term, $g$ is the contact interaction strength and $V_\mathrm{ext}$ is the externally applied potential. A similar effective field theory has been proposed and discussed to model the BEC-BCS crossover in \cite{Tempere2014,Tempere2015,Tempere2016,Tempere2018}. We include a Lagrange multiplier $\mu_\mathrm{Q}$ to fix the Klein-Gordon charge \eqref{charge}, see below. By setting $K_2=0$ and $K_1=1$ and  $\mu_\mathrm{Q}=0$, we recover the GP equation
\begin{equation}
\label{GPE}
i \partial_t \psi(\bold{x},t)=\frac{\nabla^2}{2m} \psi(\bold{x},t)+V\left(|\psi|^2\right)\psi(\bold{x},t)~,
\end{equation} 
where $V(|\psi(\bold{x},t) |^2)=\mu-g|\psi(\bold{x},t)|^2+V_\mathrm{ext}(\bold{x})$. We refer to a condensate described by the GP equation as a GP fluid. This equation is manifestly not particle-hole symmetric under the exchange $\psi \leftrightarrow \bar{\psi}$. On the other hand, particle-hole symmetry is fulfilled in the action \eqref{action} by setting $K_1=0$ and $K_2 \neq 0$.  We introduce a dimensionless representation via $\psi = \tilde{\psi} / \xi$, $\nabla = \tilde{\nabla}/\xi$, $\partial_t = c_s/ \xi \tilde{\partial}_t$ and $V = \mu \tilde{V}$, where $\xi$ is the healing length of the fluid and $c_s$ the speed of sound. This leads to the modified nonlinear Klein-Gordon (NLKG) equation
\begin{equation}
\partial^2_{\tilde{t}}  \tilde{\psi}(\bold{x},t)=\tilde{\nabla}^2 \tilde{\psi}(\bold{x},t)+\tilde{V}(|\tilde{\psi}|^2)\tilde{\psi}(\bold{x},t)+i \mu_\mathrm{Q}\partial_{\tilde{t}}  \tilde{\psi}(\bold{x},t)~.
\end{equation}
We refer to condensates evolving according to the NLKG equation as Klein-Gordon (KG) fluids. In the following we drop the tilde. We trap the fluid using a box potential of the form
\begin{align}
\label{eq:box}
V_\mathrm{ext}(\bold{x})=V_0\left(1  +\tanh\left( (|\bold{x}|-r_0)/\xi\right) \right)~. 
\end{align}
 We note that this model is a relativistic BEC \cite{Fagnocchi_2010,Haber,Grether} and a similar equation has been proposed to model cold dark matter \cite{HUANG_2012,Maga,Xiong_2014} and relativistic boson stars \cite{Monica,Chavanis_2012,Surez_2015}. \\ \indent 
In the following, we show the influence of particle-hole symmetry on the dynamics of topological defects. For the KG fluid, we introduce the canonical momentum $\Pi(x,t)=\partial_t \bar{\psi}(x,t)+i\mu_\mathrm{Q} \bar{\psi}(x,t)$ to obtain two coupled first order partial differential equations
\begin{figure}[b!]
\label{Fig_magnus}
\centering
\includegraphics[scale=0.22]{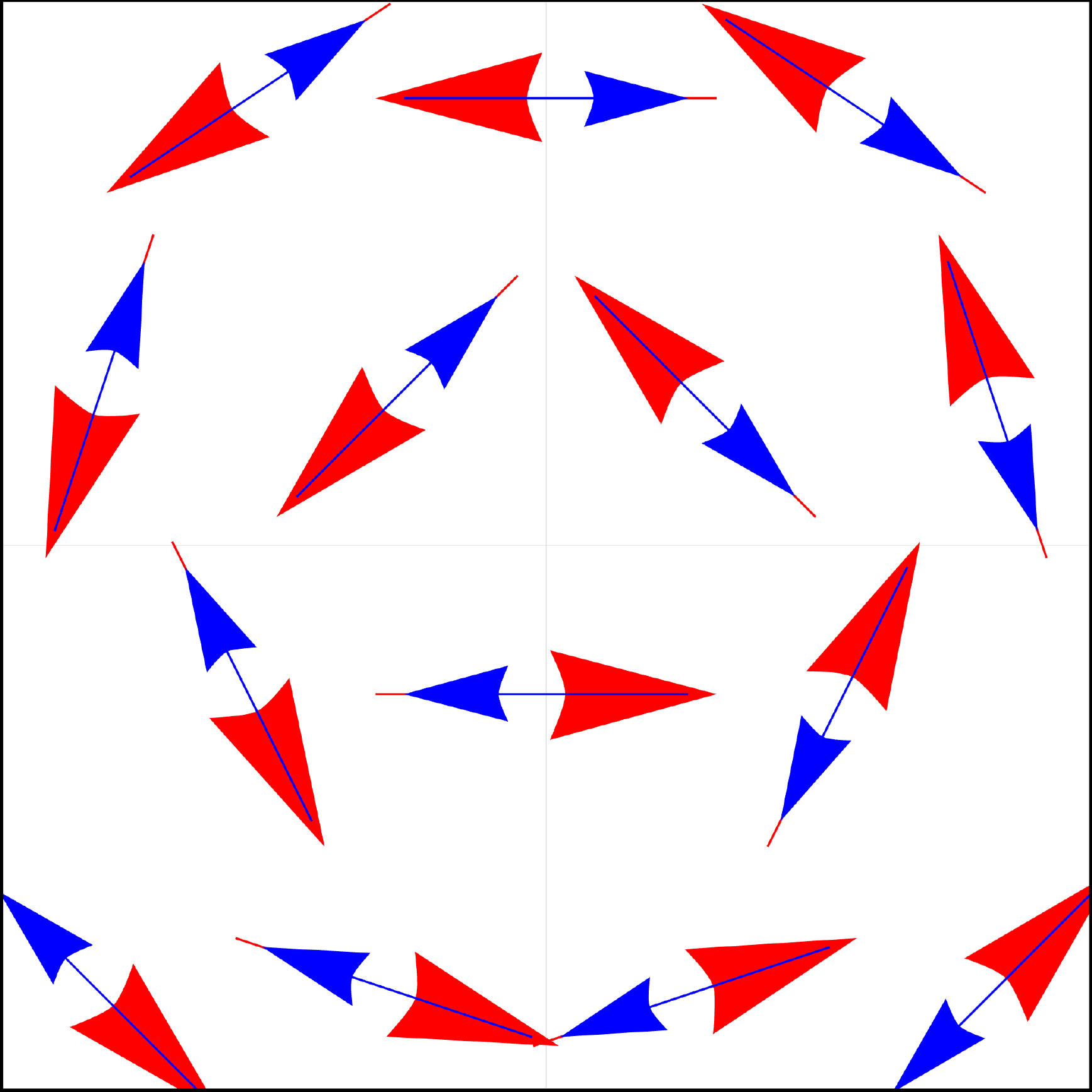}
\hspace*{0.05cm}\includegraphics[scale=0.22]{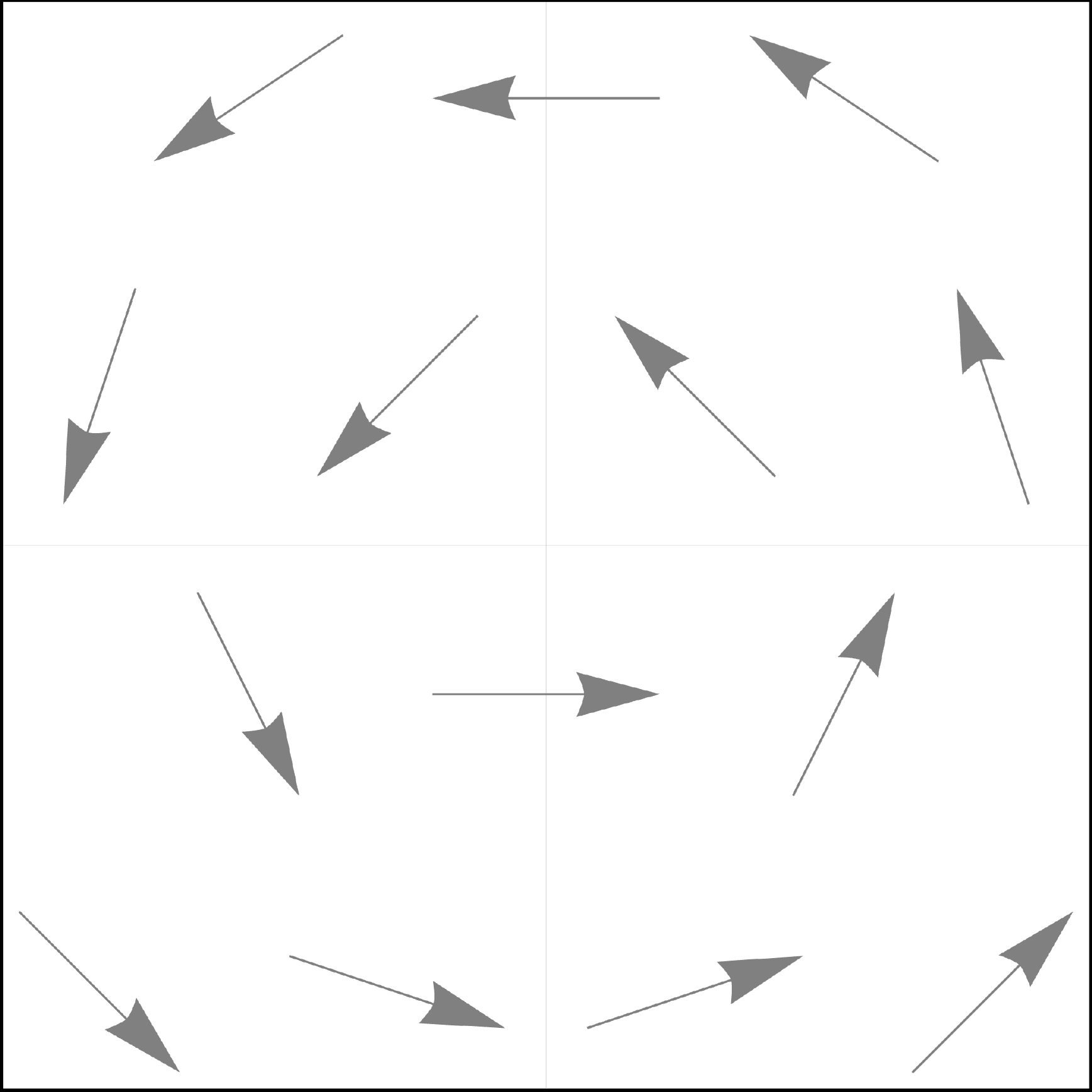} \\ \vspace*{0.15cm}
\includegraphics[scale=0.22]{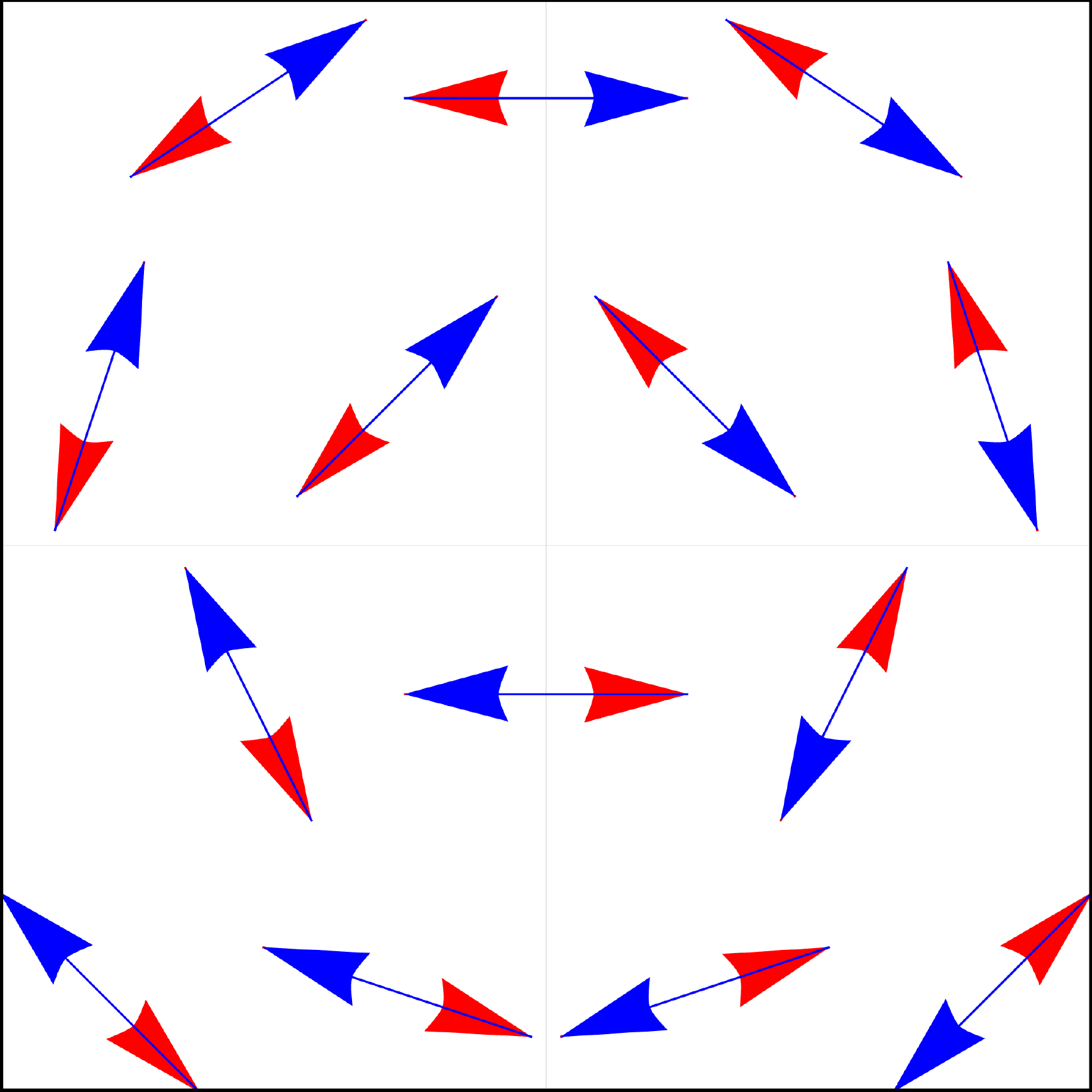} 
\hspace*{0.05cm}\includegraphics[scale=0.22]{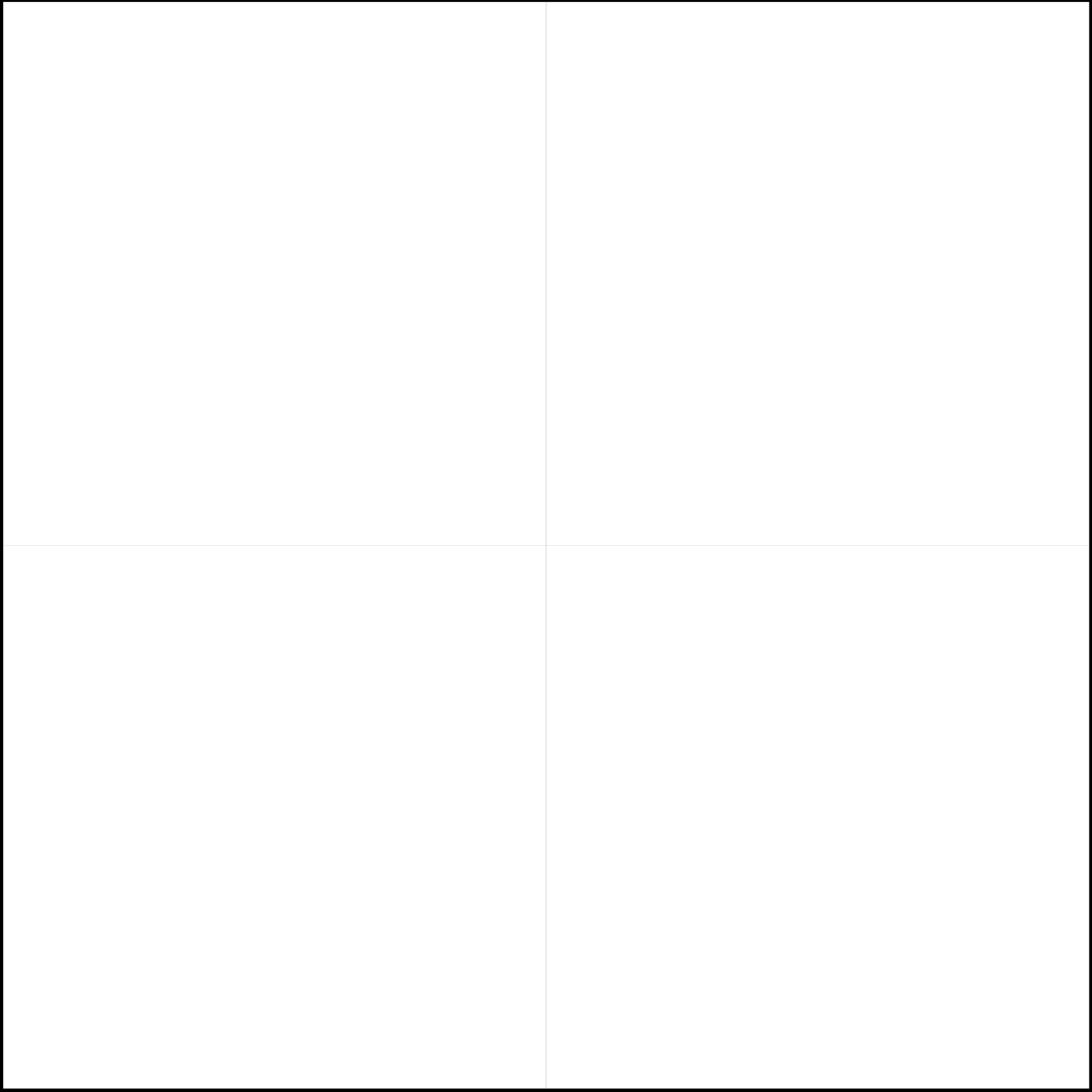} 
\begin{picture}(380,5)
\put(-20,-20){\makebox(70,504){\color{black}(a)}}
\put(-20,-20){\makebox(301,504){\color{black}(b)}}
\put(-20,-20){\makebox(70,272){\color{black}(c)}}
\put(-20,-20){\makebox(301,272){\color{black}(d)}}
\end{picture}
\label{Fig_magnus}
\caption{Schematic sketch of the local velocity fields of the particles (blue) and antiparticles (red) in (a) and (c) and the resulting total local velocity field for the $\psi$ field (gray) in (b) and (d). A unbalanced mixture of particles and antiparticles with a finite charge $Q$ (a) leads to a non-zero effective velocity field for the $\psi$ field (b). A balanced mixture of particles and antiparticles with a vanishing charge $Q$ (c) leads to a vanishing effective velocity field for the $\psi$ field (d)  }
\label{Fig_magnus}
\end{figure}
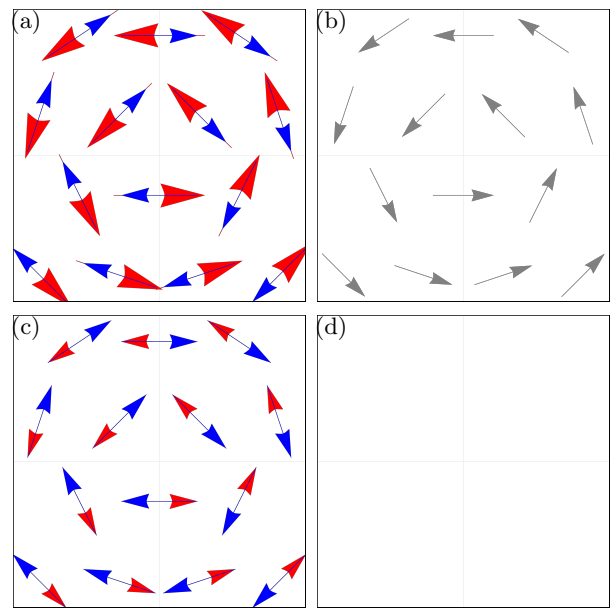 
\begin{align}
\label{KG_1}
\partial_t \psi(\bold{x},t) &=\bar{\Pi}(\bold{x},t)+i \mu_\mathrm{Q} \psi(\bold{x},t)~, \\
\label{KG_2}
\partial_t \Pi(\bold{x},t) &= \nabla^2 \bar{\psi}(\bold{x},t)+V\left(|\psi|^2\right)\bar{\psi}(\bold{x},t)-i\mu_\mathrm{Q}\Pi(\bold{x},t)~.
\end{align}
A crucial feature of a KG fluid is that the particle number $N= \int |\psi(\bold{x},t)|^2 dx$ is not conserved, in contrast to a GP fluid. Instead, in the KG fluid, the Noether charge 
\begin{equation}
\label{charge}
Q =- i  \int \left(\bar{\Pi}(\bold{x},t)\bar{\psi}(\bold{x},t)-\Pi(\bold{x},t)\psi(\bold{x},t) \right) d^2x
\end{equation} 
is conserved. The Noether charge $Q$ can be thought of as the difference of particles and holes in the system. That is, a zero Noether charge describes the situation with an equal number of particles and holes. An intuitive example for illustrating the Noether charge is a system of interacting bosons in an optical lattice with unit filling. An excitation corresponds to exciting one atom out of the lattice side and leaving behind a hole. Thus, the Noether charge stays unchanged as the same number of particles and hole where created. Another possible excitation is to excite the atom out of the lattice and further removing it from the system, which leaves a hole behind. The system then slightly goes away from unit filling as there is now an imbalance between the number of holes and particles and this corresponds to an effective nonzero Noether charge.
Another example can be envisioned in the BCS regime for nonzero temperature. Here, a RF "knife" can be used to remove some of the atoms occupying the Bogoliubov modes leading to an imbalance between particle and hole excitations. \\ \indent We apply the Madelung transformation to the field and the canonical momentum, in which the field $\psi$ is written in an amplitude-phase representation:
\begin{align}
\label{KG_phase1}
\psi(\bold{x},t)&= A(\bold{x},t)\exp(i\theta(\bold{x},t)) \\
\label{KG_phase2}
\Pi(\bold{x},t)&=\left( \frac{\dot{A}(\bold{x},t)}{A(\bold{x},t)}+i\left[ \mu_\mathrm{Q}-\dot{\theta}(\bold{x},t) \right] \right)\psi(\bold{x},t)
\end{align}
and obtain the continuity equation and particle-hole symmetric Euler equation
\begin{align}
\partial_t \rho_\mathrm{KG}+\frac{\mu_\mathrm{Q}}{2}\partial_t \rho_\mathrm{S}&=-\nabla\left( \rho_\mathrm{S} \bold{u} \right)~, \\
\label{eq:euler}
 \left( \frac{\rho_\mathrm{KG}}{\rho_\mathrm{S}}+\frac{\mu_\mathrm{Q}}{2}\right)\partial_t \bold{u} &=  \bold{u} \nabla \bold{u} +\frac{\nabla \rho_\mathrm{S}}{2\rho_0}-\frac{\nabla}{2}\left(\frac{ \Box \sqrt{\rho_\mathrm{S}}}{\sqrt{\rho_\mathrm{S}}}\right)~,
\end{align}
where we introduce the GP density $\rho_\mathrm{S}=A^2$, the KG density $\rho_\mathrm{KG}=A^2\partial_t \theta$, the velocity $\bold{u}=\nabla \theta$, and the box operator $\Box=\partial_t^2-\nabla^2$. In this representation, the charge simplifies to $Q=\int  \rho_\mathrm{KG} dx$. In the particle-hole symmetric Euler equations there is a prefactor $ \rho_\mathrm{KG}/\rho_\mathrm{S}$ in front of the time derivative of the velocity field $\partial_t \bold{u}$. This prefactor depends on the charge $Q$. This is a crucial difference to the GP Euler equation where this prefactor is always $1$. \\ \indent The particle-hole symmetric Euler equation, Eq.~\eqref{eq:euler},
 has two quantum pressure terms. One term is due to the kinetic energy of the condensate, and is proportional to $ \frac{\nabla^2 \sqrt{\rho_\mathrm{S}}}{\sqrt{\rho_\mathrm{S}}}$. It is the zero point motion of the condensate and becomes dominant if the condensate has spatial variations on short length scales \cite{pethick_smith_2008}. The second is proportional to $\frac{\partial_t^2 \sqrt{\rho_\mathrm{S}}}{\sqrt{\rho_\mathrm{S}}}$ and originates from the second-order time derivative. It only exists for particle-hole symmetric condensates.\\ \indent We present the local velocity field around a single vortex. Therefore we transform into the Feshbach-Villars basis, which translates the NLKG to coupled GP equations for the particles and antiparticles, respectively \cite{Feshbach}
\begin{align}
\psi &= \frac{1}{\sqrt{2}}(\psi^\mathrm{p}+\psi^\mathrm{a})~, \\
\Pi &= \frac{i}{\sqrt{2}}(\psi^\mathrm{a}-\psi^\mathrm{p})~.
\end{align}
\begin{figure}[b!]
\label{Fig_vort}
\centering
\includegraphics[scale=0.242]{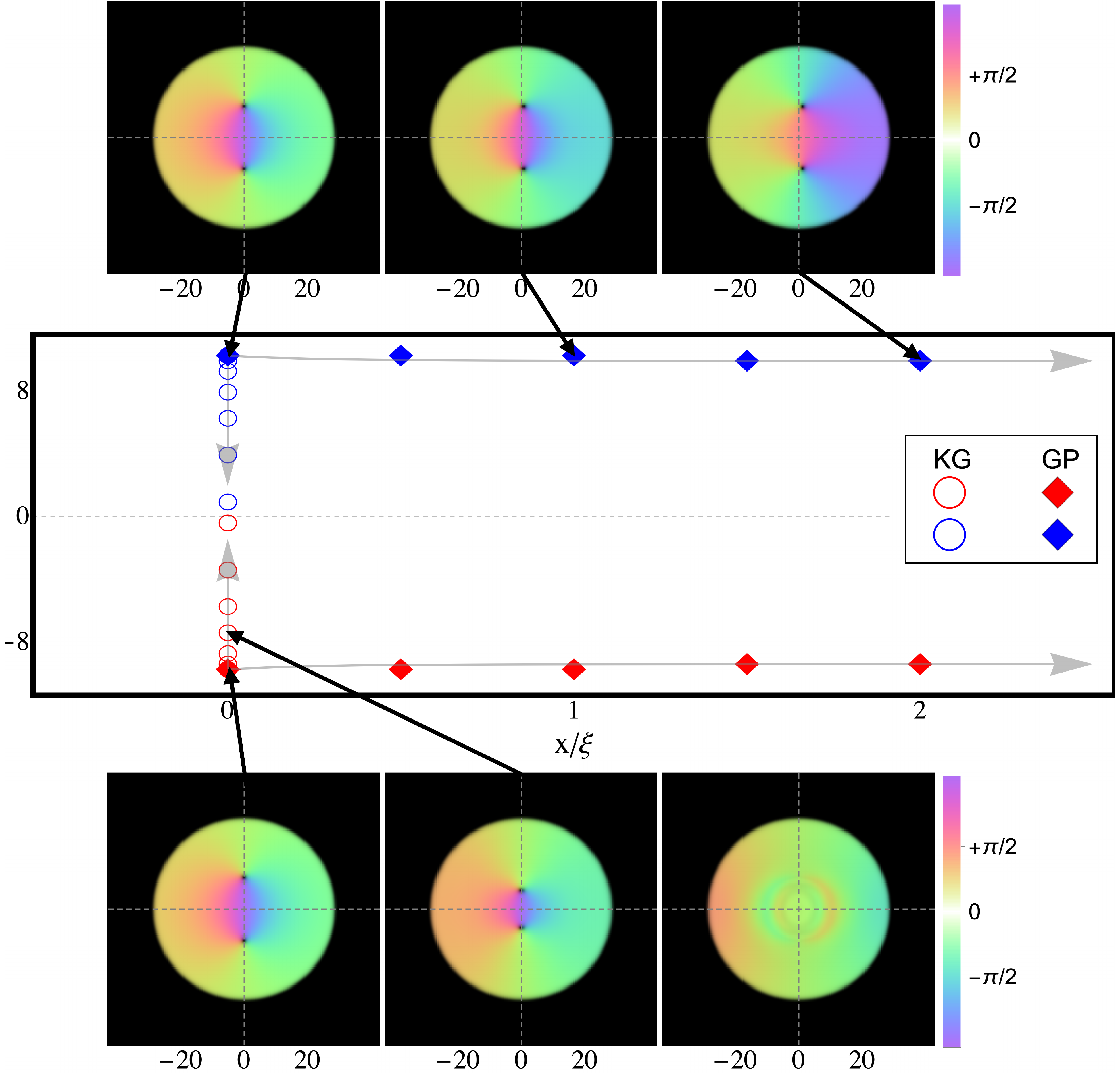}
\begin{picture}(0,5)
\put(-20,-20){\makebox(-210,604){\color{black}(a)}}
\put(-20,-20){\makebox(-210,434){\color{black}(b)}}
\put(-20,-20){\makebox(-210,225){\color{black}(c)}}
\put(-20,-20){\makebox(-180,520){\color{black}\scriptsize{\rotatebox{90}{y/$\xi$}}}}
\put(-20,-20){\makebox(-180,140){\color{black}\scriptsize{\rotatebox{90}{y/$\xi$}}}}
\put(-20,-20){\makebox(-210,330){\color{black}\scriptsize{\rotatebox{90}{y/$\xi$}}}}
\put(-20,-20){\makebox(-160,515){\color{black}\scriptsize{\rotatebox{90}{-20 \ \  \ 0 \ \  \ 20}}}}
\put(-20,-20){\makebox(-160,135){\color{black}\scriptsize{\rotatebox{90}{-20 \ \  \ 0 \ \  \ 20}}}}
\put(-20,-20){\makebox(-78,430){\color{black}\scriptsize{{x/$\xi$}}}}
\put(-20,-20){\makebox(52,430){\color{black}\scriptsize{{x/$\xi$}}}}
\put(-20,-20){\makebox(185,430){\color{black}\scriptsize{{x/$\xi$}}}}
\put(-20,-20){\makebox(52,50){\color{black}\scriptsize{{x/$\xi$}}}}
\put(-20,-20){\makebox(185,50){\color{black}\scriptsize{{x/$\xi$}}}}
\put(-20,-20){\makebox(-78,50){\color{black}\scriptsize{{x/$\xi$}}}}
\put(-20,-20){\makebox(315,140){\color{black}\large{\rotatebox{270}{KG}}}}
\put(-20,-20){\makebox(315,520){\color{black}\large{\rotatebox{270}{GP}}}}
\end{picture}
\label{Fig_vort}
\caption{Dynamics of vortex dipole pairs in a GP and KG fluid. (a) and (c) display the phase and the density of the GP fluid (a), and the KG fluid (c). (b) The circles display the location of the vortices and anti-vortices in red and blue, respectively, of the GP fluid (diamonds) and the KG fluid (circles), and difference times. The snapshots of (a) and (c) are indicated via black arrows. The gray arrows indicate the movement of the vortices in time. }
\label{Fig_vort}
\end{figure}
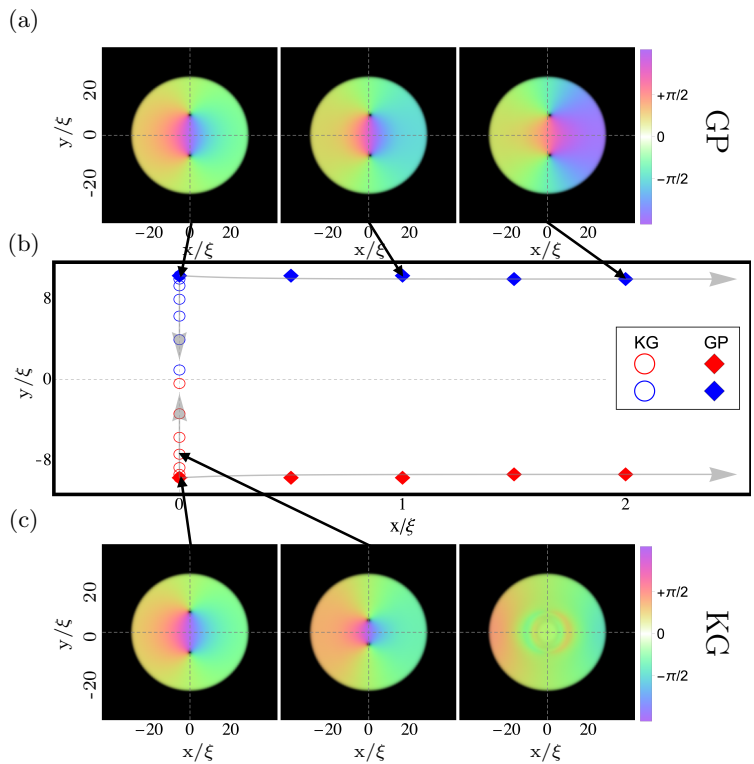  
 Next, we expand the field around the vortex core position $r_0$ with the amplitude $A^\mathrm{i}$ and phase $\theta^\mathrm{i}$ (see Eq.~\ref{KG_phase1},\ref{KG_phase2}) and propagate the location of the vortex core using the equations of motion and compare the new location with the previous location to obtain the local velocity field. See for a detailed discussion and derivation \cite{Schroeder, Groszek,supp}. For the two velocity fields we obtain
  \begin{align}
\label{mag_gp}
v^\mathrm{a} &= -\frac{(-i,1)^\mathrm{T} \cdot \vec{\nabla}(A^\mathrm{p}+A^\mathrm{a})  + (A^\mathrm{p}+A^\mathrm{a})~(1,i)^\mathrm{T} \cdot \vec{\nabla}  \theta}{A^\mathrm{a}}~, \\
v^\mathrm{p} &= \frac{(-i,1)^\mathrm{T} \cdot \vec{\nabla}(A^\mathrm{p}+A^\mathrm{a})  + (A^\mathrm{p}+A^\mathrm{a})~(1,i)^\mathrm{T} \cdot \vec{\nabla}  \theta}{A^\mathrm{p}}~,
\end{align}
where the spatial plane $(x,y)$ is represented as the complex plane $z = x + i y$. Translating this back into the
 \newpage
\begin{widetext}
\begin{figure*}[!ht]
\label{fig:2}
\centering
\begin{picture}(380,5)
\put(-53,0){\line(1,0){80}}
\put(27,-4){\line(0,1){8}}
\put(37,-4){\line(0,1){8}}
\put(37,0){\line(1,0){80}}
\put(117,-4){\line(0,1){8}}
\put(127,-4){\line(0,1){8}}
\put(127,0){\line(1,0){75}}
\put(202,-4){\line(0,1){8}}
\put(212,-4){\line(0,1){8}}
\put(212,0){\line(1,0){78}}
\put(290,-4){\line(0,1){8}}
\put(300,-4){\line(0,1){8}}
\put(300,0){\vector(1,0){90}}
\put(32,0){\makebox(0,0){..}}
\put(122,0){\makebox(0,0){..}}
\put(207,0){\makebox(0,0){..}}
\put(295,0){\makebox(0,0){..}}
\put(-10,8){\makebox(0,0){$t_1$}}
\put(75,8){\makebox(0,0){$t_2$}}
\put(165,8){\makebox(0,0){$t_3$}}
\put(254,8){\makebox(0,0){$t_4$}}
\put(339,8){\makebox(0,0){$t_5$}}
\put(-20,-20){\makebox(0,0){\color{white}(a)}}
\put(404,0){\makebox(0,0){$t / \tau$}}
\put(450,-50){\makebox(0,0){\rotatebox{270}{\large{GP}}}}
\put(450,-140){\makebox(0,0){\rotatebox{270}{\large{KG}, $Q\neq 0$}}}
\put(450,-230){\makebox(0,0){\rotatebox{270}{\large{KG}, $Q= 0$}}}
\end{picture} \\
\centering
\includegraphics[scale=1]{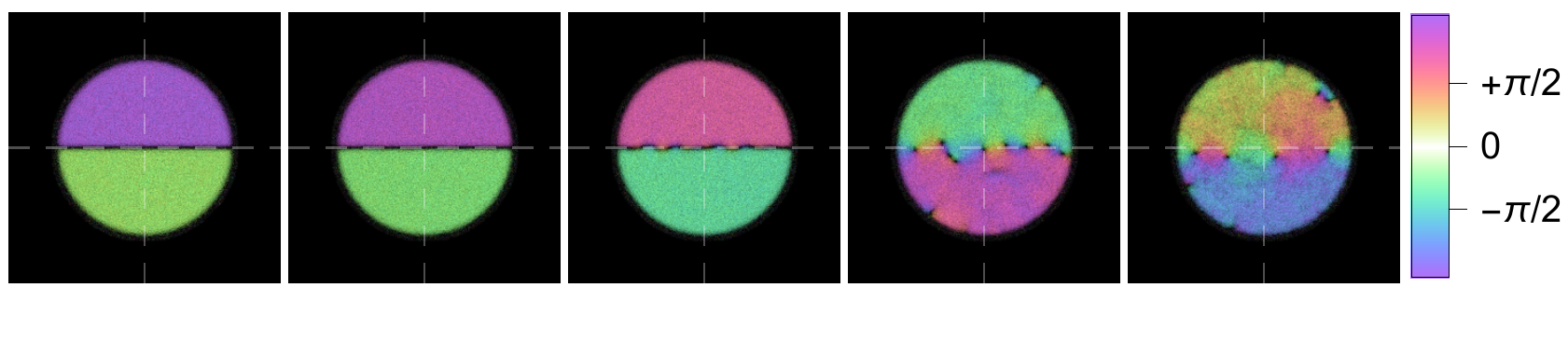}\vspace*{-0.7cm} 
\includegraphics[scale=1]{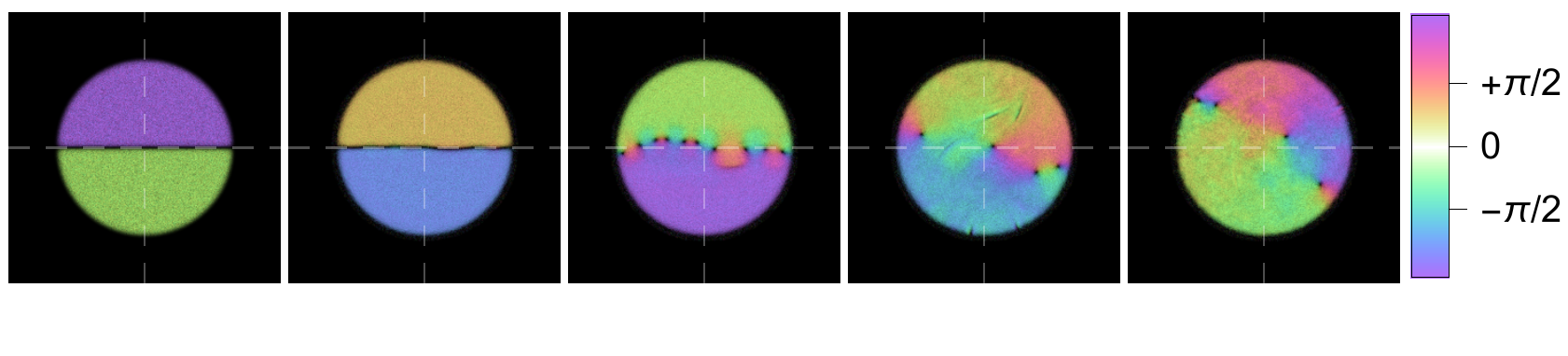}\vspace*{-0.7cm} 
\includegraphics[scale=1]{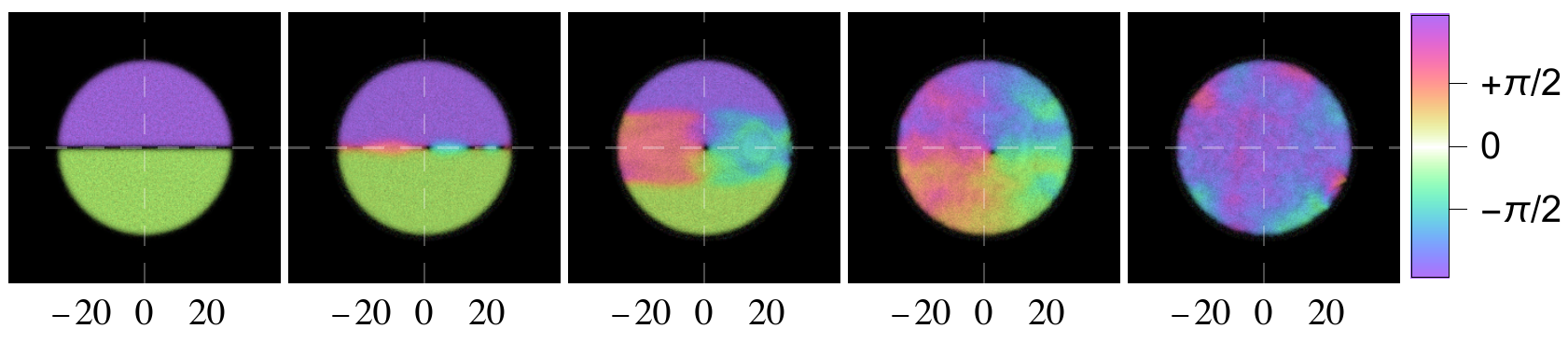} 
\begin{picture}(380,5)
\put(-35,-25){\makebox(-8,640){\color{white}(a)}}
\put(-35,-25){\makebox(166,640){\color{white}(b)}}
\put(-35,-25){\makebox(340,640){\color{white}(c)}}
\put(-35,-25){\makebox(514,640){\color{white}(d)}}
\put(-35,-25){\makebox(688,640){\color{white}(e)}}
\put(-35,-22){\makebox(-8,450){\color{white}(f)}}
\put(-35,-22){\makebox(166,450){\color{white}(g)}}
\put(-35,-22){\makebox(340,450){\color{white}(h)}}
\put(-35,-22){\makebox(514,450){\color{white}(i)}}
\put(-35,-22){\makebox(688,450){\color{white}(j)}}
\put(-35,-22){\makebox(-8,264){\color{white}(k)}}
\put(-35,-22){\makebox(166,264){\color{white}(l)}}
\put(-35,-22){\makebox(340,264){\color{white}(m)}}
\put(-35,-22){\makebox(514,264){\color{white}(n)}}
\put(-35,-22){\makebox(688,264){\color{white}(o)}}
\put(-20,-20){\makebox(-75,188){\color{black}\large{\rotatebox{90}{-20 \ 0  \ 20}}}}
\put(-20,-20){\makebox(-75,369){\color{black}\large{\rotatebox{90}{-20 \ 0  \ 20}}}}
\put(-20,-20){\makebox(-75,552){\color{black}\large{\rotatebox{90}{-20 \ 0  \ 20}}}}
\put(-20,-20){\makebox(-105,552){\color{black}\large{\rotatebox{90}{$x/\xi$}}}}
\put(-20,-20){\makebox(-105,369){\color{black}\large{\rotatebox{90}{$x/\xi$}}}}
\put(-20,-20){\makebox(-105,188){\color{black}\large{\rotatebox{90}{$x/\xi$}}}}
\put(-20,-20){\makebox(28,60){\color{black}\large{$y/\xi$}}}
\put(-20,-20){\makebox(200,60){\color{black}\large{$y/\xi$}}}
\put(-20,-20){\makebox(378,60){\color{black}\large{$y/\xi$}}}
\put(-20,-20){\makebox(550,60){\color{black}\large{$y/\xi$}}}
\put(-20,-20){\makebox(720,60){\color{black}\large{$y/\xi$}}}
\end{picture}
\label{fig:2}
\caption{Overview of the soliton instability for the three distinct cases corresponding to GP (a)-(e), KG with $Q\neq0$ (f)-(j) and KG with $Q=0$ (k)-(o). For better comparison between GP and KG the $\psi$ field density is normalised such that the maximum value is set to unity in each snapshot. The shading of the plots ranging from black to white visualises the magnitude of the field $|\psi|$, while the colormap indicates the phase. (a),(f),(k) Initial soliton seeded with white noise. (b),(g),(l) Soliton bending. (c),(h),(m) Vortices appearing after the soliton decay. The long-time dynamics of the vortices inside the trap are presented in (d),(i),(n) and (e),(j),(o). White dashed lines indicate the soliton axis and the perpendicular axis. The spatial length is expressed in terms of the healing length $\xi$. A movie showing this dynamics is presented in the supplemental material \cite{supp}.}
\label{fig:2}
\end{figure*}
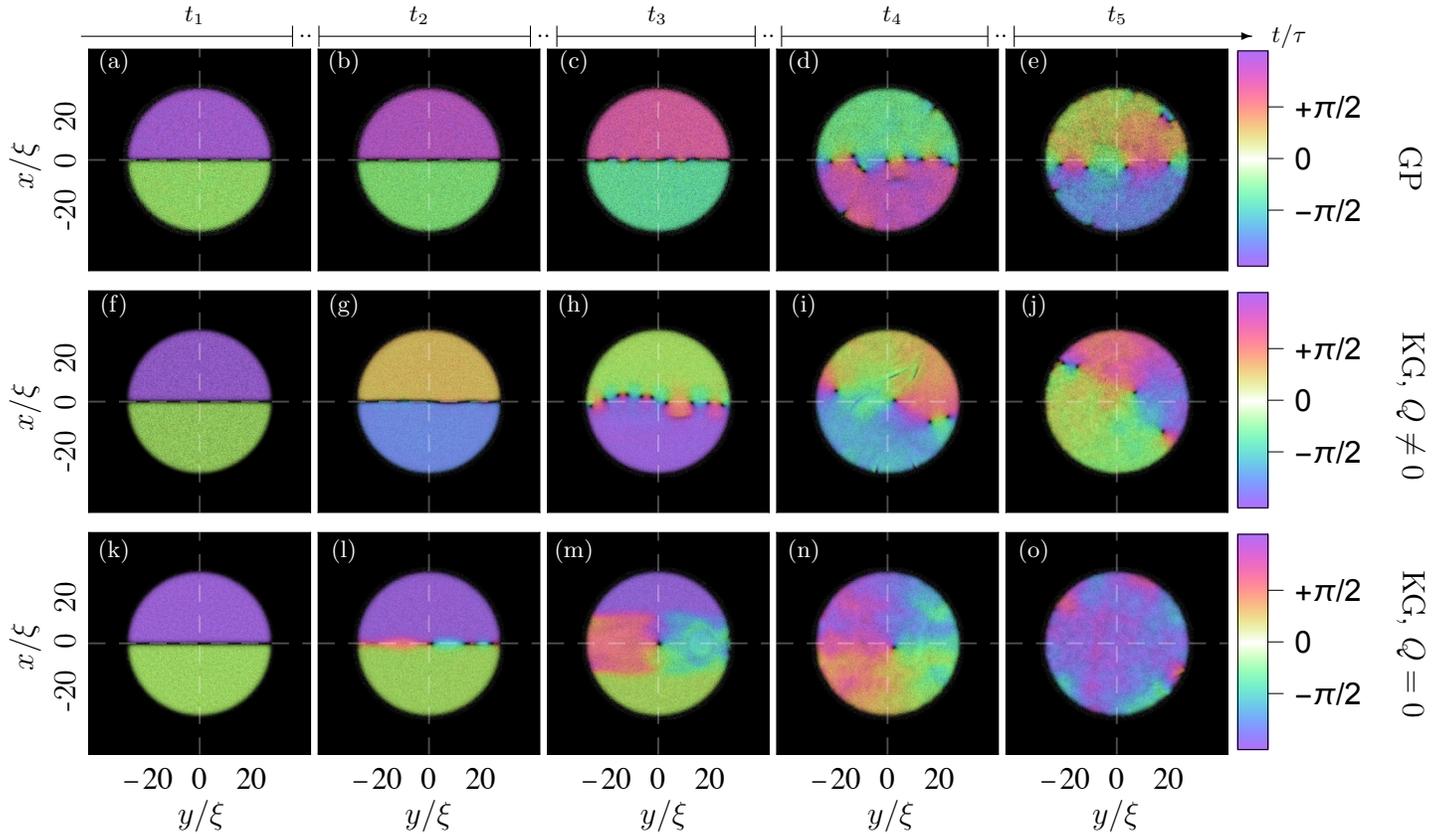
\end{widetext} 

 ($\psi$, $\Pi$) basis we obtain
\begin{equation}
\label{v_part_anti}
v^\psi =\frac{1}{\sqrt{2}}\left(v^\mathrm{p} +v^\mathrm{a} \right) =  \sqrt{2}\left(1-\frac{A^\mathrm{a}}{A^\mathrm{p}} \right) v^\mathrm{p}~.
\end{equation}

For $Q \neq 0$ we have $A^\mathrm{a} \neq A^\mathrm{p}$, which means that we obtain a non-zero velocity field. In this case the velocity is proportional to the velocity obtained for GP fluids \cite{Groszek}. For  $Q=0$, we have $A^\mathrm{p}=A^\mathrm{a}$ and $N^\mathrm{p}=N^\mathrm{a}$ with $N^i$ the total number of particles/antiparticles. For this balanced scenario the local velocity field vanishes precisely as shown in Fig.~\ref{Fig_magnus}. As pointed out before and as can be seen from Eq.~\eqref{v_part_anti}, for a finite charge $Q$ corresponding to an imbalance between particles and antiparticles the magnitudes of the velocity fields are different, see Fig.~\ref{Fig_magnus}(a), which results in a non-zero velocity field for the KG fluid $\psi$ (see Fig.~\ref{Fig_magnus}(b)). In contrast, for a balanced mixture the local velocity fields magnitudes are the same (see Fig.~\ref{Fig_magnus}(c)) and due to the opposite direction of the velocity fields the velocity field of the KG fluid vanishes (see Fig.~\ref{Fig_magnus}(d)).
To expand on our analytical predictions and to propose an experimental setup to detect vortex dynamics of KG fluids,  we simulate the equations using the pseudo-spectral method \cite{Bao_2003} for both the GP and KG fluids. We set the ratio between the chemical potential $\mu$ and the contact interaction $g$ to $\mu/g =10/\xi^2$. In the following we express all length scales in units of $\xi$. Our simulations are discretized  in a $256\times256$ grid. We choose $r_0/\xi=25$, where  $r_0$ is half of the box size, as defined in Eq.~\ref{eq:box} and resolve $\xi$ with 3 grid points.
The phase and density distribution for snapshots in real time are shown for a GP fluid (see Fig.~\ref{Fig_vort}(a)) and for a KG fluid with vanishing charge (see Fig.~\ref{Fig_vort}(c)). Circles (KG fluid with $Q=0$) and diamonds (GP fluid) in red correspond to a phase winding of $+1$, while blue corresponds to $-1$ (see Fig.~\ref{Fig_vort}(b)). The gray arrows show the flow of time in the figure. It can be seen that for a dipole distance $d_{12} > 2 \xi$ in the GP fluid the dipole will start to propel forwards perpendicular to the dipole axis and will not annihilate. In contrast, the
 KG vortex dipole will move along the dipole axis and annihilate each other, due to the absence of a velocity field. Related observations of vortex dynamics were reported in \cite{kwon2021sound}. We note that the particle-hole symmetry is the origin of this qualitatively distinct behaviour from GP fluid dynamics. We propose that the data from a future experimental realization of our proposal could be used to numerically fit the ratio of $K_1$ and $K_2$ for different interaction strengths. This links our proposal to the parameters used in the universal effective action of such systems. \\ 
 To investigate the influence of the particle-hole symmetry on the soliton dynamics, we initialise the condensate with a modified Thomas-Fermi profile \cite{pethick_smith_2008}, as described in the supplemental material \cite{supp}, for a box potential in Eq.~\eqref{eq:box} with $V_0 =10$ and $r_0 =30$ for the GP (KG) fluid and start with a soliton imprinted in the fluid. We let the condensate relax using the imaginary time propagation \cite{Barenghi_2016} extended to particle-hole symmetric fluids \cite{supp}. For the KG fluids, we set the initial canonical momentum as $\Pi=i \frac{\mu_\mathrm{Q}}{2} \psi$ with $\mu_\mathrm{Q} \in \mathbb{R}$, resulting in a charge of $Q=\mu_Q N$. Furthermore, we add $1 \%$ white noise on the initial condensate density to study the stability of solitons. \\ \indent 
The system is propagated in time according to Eqs.~\eqref{GPE}, \eqref{KG_1} and \eqref{KG_2}.
At lowest order, the Higgs mode and the Goldstone mode decouple in a particle-hole symmetric theory \cite{Pekker_2015}. Within this approximation, this initial state only induces dynamics of the Goldstone mode. However, this approximation fails in soliton and vortex solutions.  For the same parameters $\mu$ and $g$, the healing length is twice as large in the KG case compared to GPE due to the difference in the prefactor of the kinetic energy.\\ \indent In Fig.~\ref{fig:2}, we present the real time dynamics of the complex field $\psi$ shaded from white to black corresponding to decreasing amplitude, i.e. black regions denote areas with vanishing $|\psi|$. The phase of the wave function is represented as color. The wave function is normalised for each snapshot such that the maximum value is set to unity to make it easier to compare GP and KG results.\\
\indent In the GP fluid, we observe the established soliton instabilty in Figs.~\ref{fig:2}(a)-\ref{fig:2}(c) \cite{Brand} and the motion of trapped vortices in Figs.~\ref{fig:2}(c)-\ref{fig:2}(e)  \cite{Sheehy,Groszek}. The vortices move towards the edge of the condensate. As they approach the edge, the experience a net force and move along the trap boundary as depicted in Figs.~\ref{fig:2}(d) and \ref{fig:2}(e) \cite{Sheehy}. The behaviour of the KG fluid with $Q\neq0$ is similar.  As displayed in Figs.~\ref{fig:2}(f)-\ref{fig:2}(j), the soliton decays into vortices, which then move around the condensate. Similar to the GP fluid, as the phase rotates in Figs.~\ref{fig:2}(f)-\ref{fig:2}(j), the vortices experience a net force leading to their motion along the trap boundary as seen in Figs.~\ref{fig:2}(i)-\ref{fig:2}(j).\\ \indent In contrast, for the KG fluid with $Q=0$, the soliton decays into vortices that are located along the soliton axis, as shown in Figs.~\ref{fig:2}(k)-\ref{fig:2}(m). Similar results have been found in \cite{Tempere2019}. Moreover, we find that the vortices are not rotating as displayed in Figs.~\ref{fig:2}(m)-\ref{fig:2}(o) , which is consistent with Eqs.~\eqref{mag_gp}, \eqref{v_part_anti} and Fig.~\ref{Fig_vort}. When the vortices reach the trap boundary, they evaporate into the thermal cloud as shown in Figs.~\ref{fig:2}(n) and \ref{fig:2}(o). \\
\indent In conclusion, we have shown that by measuring the density profile of a 2D condensate after imprinting a soliton in a particle-hole symmetric superfluid, such as a BCS state of neutral particles, it is possible to test the effective low-energy theoretical description of the system. We have shown analytically and numerically  that for particle-hole symmetric superfluids with vanishing Noether charge the Magnus force is absent. This allows for a dipole pair of vortices to approach each other without transverse motion and to annihilate, reminiscent of recent observation in \cite{kwon2021sound}. Another consequence of the vanishing Magnus force is that a soliton does not bend as it decays into vortices. Probing these effects experimentally will reveal how well particle-hole symmetry is realised in the dynamics of superfluids or whether the non-particle-hole symmetric term, the first-order derivative in time, is the dominant contribution in the effective theory. This is crucial in understanding the notion of turbulence in particle-hole symmetric fluids such as superconductors. Our work reveals that turbulence in a BCS superconductor and its scaling laws might deviate from Kolmogorow scaling laws \cite{Kolmogorov}, which apply to classical systems as well to GP fluids. We note that our predictions could be experimentally confirmed using refined experimental technique, such as \textit{in situ} observations of two dimensional Fermi liquids when probing the BEC-BCS crossover in neutral atoms \cite{Hueck, sobirey2020observation} or the well controlled imprinting of vortex dipole pairs \cite{kwon2021sound}.
\begin{acknowledgments}	
We thank Guido Homann and Antonio Mu{\~n}oz Mateo for fruitful discussions.
This work is supported by the Deutsche Forschungsgemeinschaft (DFG) in the framework of SFB 925, Project No. 170620586, and the Cluster of Excellence ``Advanced Imaging of Matter" (EXC 2056), Project No. 390715994. J.S. acknowledges support from the German Academic Scholarship Foundation.	
\end{acknowledgments}
\bibliography{bib_main}
\end{document}


\title{Supplemental Material for\\Vortex and soliton dynamics in particle-hole symmetric superfluids}

\author{Jim Skulte}
\affiliation{Zentrum f\"ur Optische Quantentechnologien and Institut f\"ur Laserphysik, 
	Universit\"at Hamburg, 22761 Hamburg, Germany}
\affiliation{The Hamburg Center for Ultrafast Imaging, Luruper Chaussee 149, 22761 Hamburg, Germany}

\author{Lukas Broers}
\affiliation{Zentrum f\"ur Optische Quantentechnologien and Institut f\"ur Laserphysik, 
	Universit\"at Hamburg, 22761 Hamburg, Germany}

\author{Jayson G. Cosme}
\affiliation{National Institute of Physics, University of the Philippines, Diliman, Quezon City 1101, Philippines}

\author{Ludwig Mathey}
\affiliation{Zentrum f\"ur Optische Quantentechnologien and Institut f\"ur Laserphysik, 
	Universit\"at Hamburg, 22761 Hamburg, Germany}
\affiliation{The Hamburg Center for Ultrafast Imaging, Luruper Chaussee 149, 22761 Hamburg, Germany}

\maketitle
\tableofcontents

\section{Feshbach-Villars formalism}
We want to rewrite the KG equations in a Hamiltonian form of
\begin{equation}
i \partial_t \Psi = H \Psi~.
\end{equation}
This can be done via the so called Feshbach-Villars formalism \cite{Feshbach}. The Klein-Gordon equation reads
\begin{equation}
\partial^2_t \psi(r,t)=\nabla^2 \psi(x,t)+V\left(|\psi|^2\right)\psi(r,t)+ i \mu_\mathrm{Q}\partial_t \psi~.
\end{equation}
The conjugate momentum is defined as $\Pi = \frac{\partial \mathcal{L}}{\partial \left(\partial_t \psi \right)}=\partial_t \bar{\psi}+i\mu_Q \bar{\psi}$. The equations of motion read
\begin{align}
\partial_t \psi&=\bar{\Pi}+i\mu_Q \psi \\
\partial_t \Pi &= \nabla^2 \bar{\psi}(x,t)+V\left(|\psi|^2\right)\bar{\psi}(r,t)- i \mu_\mathrm{Q} \Pi.
\end{align}
The Feshbach-Villars particles or also sometimes called the particles $\psi^\mathrm{p}$ and antiparticles/holes $\psi^\mathrm{a}$ are introduced as the following linear combination of the conjugate momentum and the KG field $\psi$ 
\begin{equation}
\psi^\mathrm{p/a}= \frac{1}{\sqrt{2}} \left( \psi \pm i \Pi \right)~.
\end{equation}
That is, we can write
\begin{align}
\psi &= \frac{1}{\sqrt{2}}(\psi^\mathrm{p}+\psi^\mathrm{a}) \\
\Pi &= \frac{i}{\sqrt{2}}(\psi^\mathrm{a}-\psi^\mathrm{p})~.
\end{align}
Note that the conserved charge $Q$ in this basis simplifies to the difference between two positive definite densities as
\begin{equation}
Q = \int d^2x \left(|\psi^\mathrm{p}|-|\psi^\mathrm{a}|^2\right)~.
\end{equation}
The equations read
\begin{align}
i \partial_t \psi^\mathrm{p} &= - \frac{\nabla^2}{2}(\psi^\mathrm{a}+\psi^\mathrm{p})- (\mu_\mathrm{Q}+\mu) \psi^\mathrm{p}-V \psi^\mathrm{p}+\frac{g}{4}|(\psi^\mathrm{a}+\psi^\mathrm{p})|^2(\psi^\mathrm{a}+\psi^\mathrm{p}) \\
i \partial_t \psi^\mathrm{a} &= + \frac{\nabla^2}{2}(\psi^\mathrm{a}+\psi^\mathrm{p})+ (\mu_\mathrm{Q}+\mu) \psi^\mathrm{a}+V \psi^\mathrm{a}-\frac{g}{4}|(\psi^\mathrm{a}+\psi^\mathrm{p})|^2(\psi^\mathrm{a}+\psi^\mathrm{p})~.
\end{align}
\section{Local velocity field}
\subsection{$\psi$ and $\Pi$ basis}
We start by denoting our field and the canonical momentum as
\begin{align}
\psi(r,t=0) &\equiv \psi_{r,0} =P\left( \vec{z}-\vec{z_0} \right)A\exp\left(i \theta \right)~, \\
\label{momentum}
\Pi(r,t=0) &\equiv \Pi_{r,0} = \bar{P}\left( \vec{z}-\vec{z_0} \right)\left(\dot{A}+i A \left[ \mu_\mathrm{Q}-\dot{\theta} \right]\right)\exp\left(-i \theta \right)
\end{align}
where $P((x,y)^\mathrm{T})=x+iy$ and the vortex core position is located at $ \vec{z}_0= (x_0,y_0)^\mathrm{T}$. Note that $A(r,t)$ and $\theta(r,t)$ are real valued and smoothly varying functions of space and time. This field features a vortex, where $P(\vec{z}-\vec{z}_0)$ captures both the phase winding and the condensate density around the vortex core.
We can propagate the field for an infinitesimally small $\Delta t$ according to the Klein-Gordon equation as:
\begin{align}
\psi_{r,\Delta_t} &\equiv \psi_{r,0} + \Delta t\left( \bar{\Pi}_{r,0}+i \mu_\mathrm{Q}\psi_{r,0}\right)~, \\
\label{small_t}
\Pi_{r,\Delta_t} &\equiv \Pi_{r,0} + \Delta t\left( \nabla^2 \bar{\psi}_{r,0}+V\left(|\psi_{r,0}|^2\right)\bar{\psi}_{r,0}-i\mu_\mathrm{Q}  \Pi_{r,0}\right)~.
\end{align}
Using  $\nabla \bar{z} = (1,-i)^\mathrm{T}= \vec{e}_x-i \vec{e}_y$ and $\nabla^2 z =0$, we obtain
\begin{align}
\nabla^2 \bar{\psi}_{r,0} =& [\bar{P}(\vec{z}-\vec{z}_0)\nabla^2 A + 2(1,-i)^\mathrm{T} \cdot \vec{\nabla} A  -2i \{ \bar{P}(\vec{z}-\vec{z}_0)\vec{\nabla} A+A~(1,-i)^\mathrm{T} \} \cdot \vec{\nabla}  \theta \notag  \\ &+\bar{P}(\vec{z}-\vec{z}_0)A \{ -i\nabla^2  \theta-(\vec{\nabla}  \theta)^2 \}]\exp( i \theta)~.
\end{align} 
The canonical momentum in Eq.~\eqref{momentum} vanishes at the vortex core position $\vec{z}_0$ at $t=0$. Thus, we can write the new vortex core position as $\vec{z}_{\Delta t}=\vec{z}_0+ \vec{v} \Delta t$ after the small time $\Delta t$, where the canonical momentum still vanishes. Evaluating $\bar{\Pi}(z_{\Delta t},\Delta t)$, and demanding it to be zero while only keeping lowest order terms in $\Delta t$, we obtain
\begin{align}
\label{delta_z}
0 \approx  &  \bar{\Delta} z \left(\dot{A}-iA( \dot{\theta}-\mu_\mathrm{Q} )\right) + \Delta t \left( 2(1,-i)^\mathrm{T} \cdot \vec{\nabla} A + 2A~(1,i)^\mathrm{T} \cdot \vec{\nabla}  \theta \right)~,
\end{align} 
where we used $\bar{P}(\vec{z}_{\Delta t}-\vec{z}_0)= \bar{\Delta} z$. We rearrange this and take the complex conjugate to obtain an expression for the velocity as
\begin{equation}
\label{mag_kg}
P(\vec{v}) \equiv \frac{\Delta z}{\Delta t}= -2\frac{(1,-i)^\mathrm{T}\cdot \vec{\nabla} A - A~(i,1)^\mathrm{T} \cdot \vec{\nabla}  \theta}{\dot{A}+iA (\dot{\theta}-\mu_\mathrm{Q})}~.
\end{equation}
Similarly, the velocity field for the GP fluid can be obtained as \cite{Schroeder, Groszek}
\begin{equation}
\label{mag_gp}
P(\vec{v}) \equiv \frac{\Delta z}{\Delta t}= (-i,1)^\mathrm{T} \cdot \vec{\nabla} A  + A~(1,i)^\mathrm{T} \cdot \vec{\nabla}  \theta~.
\end{equation}
The discussion changes if we choose  $\Pi=0$, ($Q=0$) as the initial state. The derivation holds up to Eq. \eqref{delta_z}.  From this equation we obtain
\begin{equation}
0 \approx  \Delta t \left( 2(1,-i)^\mathrm{T} \cdot \vec{\nabla} A + 2A~(1,i)^\mathrm{T} \cdot \vec{\nabla}  \theta \right)~.
\end{equation} 
As $\Delta t \neq 0$ we deduce that 
\begin{equation}
\left( 2(1,-i)^\mathrm{T} \cdot \vec{\nabla} A + 2A~(1,i)^\mathrm{T} \cdot \vec{\nabla}  \theta \right) =0~.
\end{equation}
Hence, there is no dynamics up to lowest order for the canonical momentum $\Pi$ and the field $\psi$ as can be seen from plugging this result into \eqref{small_t}.
\subsection{Particle and antiparticle basis}
In the following we will redo the steps as before but now using the particle and antiparticle fields to obtain a physical explanation of the vanishing velocity field for the zero charge case. \\
We denote our fields as
\begin{align}
\psi^{\mathrm{a/p}}(r,t=0) &\equiv \psi^{\mathrm{a/p}}_{r,0} =P\left( \vec{z}-\vec{z_0} \right)A^{\mathrm{a/p}}\exp\left(i \theta \right)~,
\end{align}
where we have assumed that the two fields stay in phase. Note that for the choice of $A^\mathrm{a}= A^\mathrm{p}$ the Noether charge is vanishing. For small times $\Delta t$ the fields can be propagated in time via
\begin{align}
\psi^\mathrm{a/p}_{r,\Delta_t} &\equiv \psi^\mathrm{a/p}_{r,0} \mp i \Delta t\left( \frac{\nabla^2}{2}(\psi^\mathrm{a}_{r,0}+\psi^\mathrm{p}_{r,0})- (\mu_\mathrm{Q}+\mu) \psi^\mathrm{a/p}_{r,0}+\frac{g}{4}|(\psi^\mathrm{a}_{r,0}+\psi^\mathrm{p}_{r,0})|^2(\psi^\mathrm{a}_{r,0}+\psi^\mathrm{p}_{r,0})\right)~.
\label{small_t_ap}
\end{align}
By performing the same steps as for the $\psi$ and $\Pi$ basis we obtain for the local velocity fields
\begin{align}
\label{mag_gp}
P(\vec{v^\mathrm{p}}) = \frac{(-i,1)^\mathrm{T} \cdot \vec{\nabla}(A^\mathrm{p}+A^\mathrm{a})  + (A^\mathrm{p}+A^\mathrm{a})~(1,i)^\mathrm{T} \cdot \vec{\nabla}  \theta}{A^\mathrm{p}}~ \\
P(\vec{v^\mathrm{a}}) = -\frac{(-i,1)^\mathrm{T} \cdot \vec{\nabla}(A^\mathrm{p}+A^\mathrm{a})  + (A^\mathrm{p}+A^\mathrm{a})~(1,i)^\mathrm{T} \cdot \vec{\nabla}  \theta}{A^\mathrm{a}}~.
\end{align}
We note that $\vec{v^\mathrm{p}}A^\mathrm{p}=-\vec{v^\mathrm{a}}A^\mathrm{a}$ and hence
\begin{equation}
\vec{v^\mathrm{p}} = -\frac{A^\mathrm{a}}{A^\mathrm{p}}\vec{v^\mathrm{a}}~.
\end{equation}
The velocity field of the $\psi$ field can be recovered by the linear superposition as
\begin{equation}
\vec{v^\psi} =\frac{1}{\sqrt{2}}\left(\vec{v^\mathrm{p}} +\vec{v^\mathrm{a}} \right) =  \sqrt{2}\left(1-\frac{A^\mathrm{a}}{A^\mathrm{p}} \right)\vec{v^\mathrm{p}}~.
\end{equation}
We can understand the vanishing local velocity of the total field $\psi$ at vanishing Noether charge $A^\mathrm{p}=A^\mathrm{a}$ as the result of two counter rotating fields consisting of the particle and antiparticle field. This results is in a similar spirit as the zero magnetization results for two component spinors in \cite{turner}.
\section{Thomas-Fermi-Approximation}
In order to find steady state solutions we demand that $\partial_t \psi =0$ and $\partial_t \Pi=0$, while neglecting the kinetic energy. This approximation is justified for large particle numbers and repulsive interactions \cite{pethick_smith_2008}. We obtain the two equations
\begin{align}
0&=\bar{\Pi}+i\mu_Q \psi \\
0 &=\left(V(x)-\mu+g|\psi|^2 \right)\bar{\psi}(r,t)- i \mu_\mathrm{Q} \Pi.
\end{align}
We find for the region, where $V(x) \leq (\mu+\mu_\mathrm{Q}^2)$
\begin{align}
|\psi(x)|^2 &= \frac{\mu+\mu_\mathrm{Q}^2-V(x)}{g} \\
\Pi(x) &= -i \mu_\mathrm{Q} \bar{\psi} (x)
\end{align}
 and $\psi=\Pi=0$ otherwise. The boundary of the fluid is at $V(x) = (\mu+\mu_\mathrm{Q}^2)$.
\section{Imaginary time method for particle-hole symmetric fluids}
The particle-hole symmetry enforces the energy spectrum to be symmetric around $E=0$. We will follow the same line of arguments as in \cite{Barenghi_2016}. For Schr\"odinger like equations the fields can be propagated via
\begin{equation}
\psi^\mathrm{p/a}(x,t+\Delta t) = e^{-i \Delta t H}\psi^\mathrm{p/a}(x,t)~.
\end{equation}
Expanding the fields in energy modes as $\psi^\mathrm{p/a}(x,t)= \sum_m a^{p/a}_m(t) \phi^{p/a}_m(x)$, we obtain
\begin{equation}
\psi^\mathrm{p/a}(x,t+\Delta t) =\sum_m a^{p/a}_m(t) \phi^{p/a}_m (x) e^{\mp i \Delta t E_m}
\end{equation}
To ensure that the wavefunction decays into the lowest energy state we rotate $\Delta t$ for the particles and antiparticles as $\Delta t \rightarrow \mp i \Delta t$. By doing so we obtain
\begin{equation}
\psi^\mathrm{p/a}(x,t+\Delta t) =\sum_m a^{p/a}_m(t) \phi^{p/a}_m (x) e^{- \Delta t E_m}~.
\end{equation}
After each propagating step we fix the total number 
\begin{equation}
N^{p/a}= \int d^2x |\psi^\mathrm{p/a}|^2~,
\end{equation}
which ensures that the total charge $Q$ stays constant.
\bibliography{biblio}